\documentclass[10pt,conference]{IEEEtran}
\IEEEoverridecommandlockouts

\usepackage{cite}
\usepackage{amsmath,amssymb,amsfonts}
\usepackage{algorithmic}
\usepackage{graphicx}
\usepackage{textcomp}
\usepackage{xcolor}
\usepackage{url}
\usepackage{balance}
\newif\ifdraft
\drafttrue 
\definecolor{colorIA}{rgb}{0.13, 0.7, 0.67}
\newcommand{\red}[1]{\textcolor{red}{#1}}
\newcommand{\mv}[1]{\ifdraft\textcolor{blue}{MV: #1\fi}}

\usepackage{ifthen}
\newboolean{blinded}
\setboolean{blinded}{false}

\newtheorem{definition}{Definition}

\title{%
Defending Event-Triggered Systems against Out-of-Envelope Environments
\ifthenelse{\boolean{blinded}}{}{\thanks{This research was funded by the Luxembourg National Research Fund (FNR) and by the German Research Council DFG through the FNR-Core-Inter grant ReSAC (C21/IS/15741419). This article appeared in the RTAutoSec Workshop organized by Monowar Hasan and Mohammad Hamad, which was co-located to ECRTS 2025. License: CC-BY-SA-4.0}.}}

\author{
  \ifthenelse{\boolean{blinded}}{
  }{
    \IEEEauthorblockN{
      Marcus V\"olp$^1$,
      Mohammad Ibrahim Alkoudsi$^2$,
      Azin Bayrami Asl$^1$,\\ 
      Kristin Kr\"uger$^2$
      J\'{u}lio Mendon\c{c}a$^1$,
      Gerhard Fohler$^2$
    }
    \IEEEauthorblockA{
      $^1$Interdisciplinary Centre for Security Reliability and Trust (SnT), University of Luxembourg\\
      e-mails: \url{marcus.voelp@uni.lu}, \url{azin.bayramiasl@uni.lu}, \url{julio.mendonca@uni.lu} \\
    }
    \IEEEauthorblockA{
      $^2$Rheinland-Pf\"alzische Technische Universit\"at Kaiserslautern - Landau\\
      e-mails: \url{alkoudsi@rptu.de}, \url{kristin.krueger@rptu.de}, \url{gerhard.fohler@rptu.de}\\
    }
  }
}

\begin{document}

\maketitle

\begin{abstract}
  
  The design of real-time systems is based on assumptions about environmental conditions in which they will operate. We call this their safe operational envelope.
  Violation of these assumptions, i.e., out-of-envelope environments, can jeopardize  timeliness and safety of real-time systems, e.g., by overwhelming them with interrupt storms. 
A long-lasting debate has been going on over which design paradigm, the time- or event-triggered, is more robust against such behavior.  
     
  In this work, we investigate the claim that  time-triggered systems are immune against out-of-envelope behavior and how event-triggered systems can be constructed to defend against being overwhelmed by interrupt showers.
  We introduce importance (independently of priority and criticality) as a means to express which tasks should still be scheduled in case environmental design assumptions cease to hold, draw parallels to mixed-criticality scheduling, and demonstrate how event-triggered systems can defend against out-of-envelope behavior.
\end{abstract}

\begin{IEEEkeywords}
time-triggered systems, event-triggered systems, out-of-envelope behavior    
\end{IEEEkeywords}

\section{Introduction}
\label{sec:introduction}

Real-time computer systems are typically designed to function within an operational envelope that dictates all possible behaviors of the environment that encloses them. 
The specification of an operational envelope allows for design-time assurance that the real-time system will behave in a timely manner if it operates within this envelope \cite{Delta4}. 

A common argument in the decade-old debate between the time- and event-triggered paradigms' experts is the question how to cope with out-of-envelope behavior.
On one side, Kopetz's tremendous achievement of predictable, fault-tolerant, composable, and modularly certifiable systems, enabled by the time-triggered architecture~\cite{TTA2003}, and on the other side, there have been various attempts~\cite{6818368,5755437,8264463} towards achieving similar properties in event-triggered systems.

In the time-triggered architecture (TTA), 
all system components share a globally synchronized sparse-time base~\cite{TTA2003, SparseTime1992}. 
All important events and activities, e.g., sampling the environment and exerting control over it, are internalized to periodically occurring time points and intervals on this time base.  
Intermediate control algorithms are also set to execute and communicate their results within predetermined intervals of sparse time (time-slots), 
where each task or message gets a single or multiple exclusive time-slots on the respective resource. 
This temporal isolation rules out interference from other tasks or messages executing in different time-slots.  

Therefore, as long as the environment matches the regime dictated by the specified sparse time-slots, and as long as the globally maintained time base remains immune to faults and attacks, we obtain the above properties and the elegance and beauty of the TTA.
However, once the environment moves beyond this envelope, the situation becomes 
more complicated.

Event-triggered systems internalize events immediately when the sensors and devices at the periphery of the system observe them. Sensors, network interfaces, and other devices are connected to interrupt lines, which, when raised, signal the processor to interrupt its current control flow and activate a corresponding service routine. 
Such routines in turn handle the interruption, internalize the event (e.g., by associating it with a timestamp) and release tasks that are triggered by this event. 
The routines then invoke the scheduler to see whether the released tasks can be scheduled immediately or if their execution needs to be deferred (e.g. because a higher priority task is currently running or because resource locking protocols (such as stack-based ceiling~\cite{128747}) demand deferred execution). 

Immediate internalization makes event-triggered systems vulnerable to interrupt storms, triggered by faulty sensors and devices, and to unanticipated events, when environmental situations are in violation of the assumed operational envelope. 
Designers of event-triggered systems typically equate both situations as undesirable behavior and construct their systems to avoid both by ignoring events that occur in addition to those anticipated.
This risks missing important deviations from the norm and taking appropriate actions to keep the system safe, even when normal operation can no longer be guaranteed.

In this work, we take a different approach and survey what it takes in event-triggered systems to defend against out-of-envelope behavior.
Our goal is to design a system that will ultimately be able to respond in a timely and predictable manner to unforeseen situations while being equipped to carefully trade-off less important tasks, and to defend itself against error situations that would overwhelm the system.

We start in Section~\ref{sec:out_of_envelope}, by giving concrete examples, to highlight what we mean by out-of-envelope behavior and discuss how contemporary time- and event-triggered systems respond to such behavior. We then introduce in Section~\ref{sec:feasibility} a new feasibility criterion, inspired by mixed-criticality scheduling~\cite{4408308,burns2013mixed}, to ensure systems continue to respond to important events, even if they can no longer sustain all of their normal behavior. 
We will also discuss why one might want to introduce importance separate from criticality.
Section~\ref{sec:defense} illustrates how event-triggered systems can defend against being overwhelmed by out-of-envelope behavior. 

Of course, we are not the first aiming to design defensive event-triggered systems and aspects will have to remain as open questions, which is why, in Section~\ref{sec:limitations} and Section~\ref{sec:related_work}, we discuss limitations and open questions and we relate our work to the works of others, before concluding our work in Section~\ref{sec:conclusions}.

\section{Out-of-Envelope Behavior}
\label{sec:out_of_envelope}

Real-time computer systems interact with their environment by sensing environmental conditions together with the observable states of the machines they control. They receive alarms from specific sensors and influence the environment by actuating parts of these machines.
Failure to do so in a correct and timely manner often has severe consequences and puts at risk the safety of the controlled machines, of the people in their proximity, or of the environment in which they operate. 

When designing real-time systems, developers define an operational envelope, in which they characterize their assumptions about the environment and specifically what events they expect and how frequently and with which separation they expect them to occur~\cite{6818368}. 
For example, when controlling a nuclear plant, sensors reveal whether valves are open or closed.
Other sensors read out the pressure and water level in the pressurizer and, more generally, dynamic processes are observed with at least double the frequency of the highest signal frequency that is to be expected.

In this work, we are primarily interested in external events, happening at the system's periphery, such as reactor pressure levels exceeding a certain threshold, which is observed by a sensor at an appropriate location. While pressure remains within certain bounds (i.e., within the envelope), the sensor-measured pressure values change only rarely and in a continuous manner. Beyond these bounds however, bubbles in the cooling water may cause fast alterations between low- and high-pressure situations at the same sensor, indicating a serious and unexpected situation, which in the case of Three-Mile Island has already led to a serious incident~\cite{three-mile-island}.

Time-triggered systems internalize these signals at pre-defined and, thanks to the globally synchronized sparse time base, globally known events.
Let $T_{\mathit{sample}}$ be the pre-defined sampling period of an observed real-time entity, e.g., a valve in the nuclear plant.  
Let $t_{\mathit{valve}}$ and $t'_{\mathit{valve}}$ denote the time at which the observed valve changes state to open and closed, respectively. 
If the duration $t'_{\mathit{valve}}-t_{\mathit{valve}}$ is smaller than $T_{\mathit{sample}}$, then the 
time-triggered system will not be able to recognize and internalize the opening event of the valve $t_{\mathit{valve}}$. 
The operational envelope defines that developers do not expect more frequent events and designing systems according to that envelope implies that such a behavior will not be considered by the developed system.


In event-triggered systems, in addition to the possibility of sampling events, changes are communicated almost immediately from the sensor to the computer system by means of raising interrupts at the latter.
Interrupts are hardware signals (delivered over dedicated lines or as PCI and then memory bus messages) to one of the processors' interrupt controllers, which in turn causes the processor to preempt its current task and enter the operating system's \emph{top-half} interrupt service routine.
Once interrupt occurrence is recorded and, for level-triggered interrupts, the source is masked, the operating system (OS) decides whether it processes the interrupt immediately or returns to the preempted task for deferred handling in a so-called \emph{bottom half}.
The critique put forward is that this recording and masking in the top half together with the kernel entry, already interferes enough with the scheduled tasks to reduce predictability and hinder independent certification.
In practice, however, as long as the environment behaves as depicted in the operational envelope, preemptions can be anticipated and accounted for in the tasks' worst-case execution times (WCETs), in particular if the handling of the remaining part of the interrupt (the bottom half) can be deferred and scheduled like regular jobs.

Out-of-envelope behavior happens when unexpected events occur. Event-triggered systems will then have more interrupts raised than anticipated or combinations of interrupts that were not considered. 
We have already seen that time-triggered systems are immune to such occasions, since they will not internalize out-of-envelope behavior, and will proceed as if none of the missed events have happened.
This is where event-triggered systems have the potential for a more attenuated response, by continuing to respond also to unforeseen events and specifically to alarms. It will even be possible to plan for responses to unlikely event combinations, without reserving time for these responses in the regular schedule.

Unfortunately, by responding to out-of-envelope behavior, event-triggered systems also make themselves vulnerable to increased interference, in case multiple events happen close together, and specifically to interrupt storms.
In fact, naive implementations, will simply crash event-triggered systems in such situations due to the kernel stack overflowing when pushing interrupted state in an uncontrolled manner. The execution of high-priority tasks may be delayed when executing the top-halves of a storm of interrupts pertaining to the execution of low-priority tasks, leading to deadline misses and risking safety.
Of course there are several works that address this concern albeit within the assumed environmental envelope~\cite{10.1145/1629395.1629419,7299849,10.1145/2220336.2220344,6257578,article,4700438}.
For instance, Parmer and West~\cite{4700438} schedule bottom halves in deferable servers to limit the budget bottom halves may consume within a given amount of time. However, they dimension the budget to consider only events that have been anticipated in the system's operational envelope. 

In this work, we also consider out-of-envelope behavior with the goal of equipping systems with the possibility to respond to important events, specifically alarms, even if they happen more frequently than anticipated.
At the same time, our goal is to defend systems against faulty sensors raising alarms continuously and, in turn, overwhelming the system. 

\section{Out-of-Envelope Feasibility}
\label{sec:feasibility}

Our goal is to allow event-triggered systems to remain responsive to out-of-envelope behavior, while defending against being overwhelmed. 
In normal situations, when the environment behaves as in the assumed envelope, all events should be internalized, and corresponding tasks released and scheduled, including the events' bottom-half handlers in case part of the immediate event response can be deferred.
Once important events occur more frequently than anticipated, but still within certain bounds, the system should obtain the possibility to trade off the handling of less important events and their corresponding tasks and still inform operators about triggered alarms and still take the actions that are required to keep the system safe.
However, when exceeding these bounds, sensors must be considered faulty and should be suppressed to these bounds, while indicating that more events occurred than could be internalized.

Since we specifically consider alarms, we assume a classical mix of sporadic and periodic tasks, as captured in the sporadic task model.
Events (including alarms or the timers firing at specific points in time) release jobs $\tau_{i,j}$ of tasks $\tau_i$ in the task set $\Gamma$.
Each task is characterized by the tuple $(C_i, D_i, T_i, I_i)$, where for simplicity we assume relative deadlines $D_i$ are implicitly defined by the task's minimum inter-arrival time $T_i$ (i.e., $D_i = T_i$).
Tasks are feasible if all jobs receive $C_i$ time between their release $r_{i,j}$ and their absolute deadline $r_{i,j} + D_i$.

In order to characterize the importance of an event, and hence of the jobs it releases, we introduce an \emph{importance} value $I_i$, which we use to create a total order of importance for each task.
In mixed-criticality systems, importance can be mapped to criticalities (e.g., by assigning low-criticality tasks an importance up to a certain level $l$ --- tasks with importance $I_i \in [0, l)$ are LO --- and high-criticality tasks a higher value) and $C_i$ and other task parameters may be criticality-level dependent (e.g., $C_i$ may be a vector $C_i(l_i)$ where $l_i$ is the criticality level of the task).
In addition, importance may be mapped to priorities unless this interferes with priority assignment (e.g., tasks with a larger period $T_i$ may be more important than those with a smaller period, while rate-monotonic scheduling would assign them lower priorities, which we would like to allow). 
However, there are also cases where importance should be handled independently of criticality and priority. 
Criticality is commonly used to grant additional resources to those tasks that are more critical to maintain the safety of the system. It does not distinguish internal and external failure modes as to why such a task would need additional resources. In our work, 
we are primarily concerned with internalizing and allowing the handling of external events, such as the frequent change of pressure levels when bubbles rise in the cooling water, to then take appropriate actions. In that sense, events arriving outside the anticipated envelope may trigger a criticality change, but not necessarily vice versa.
Therefore, in the above model, in normal situations, all events should be internalized and all tasks be scheduled, irrespective of their importance.

To capture out-of-envelope behavior, we introduce as additional parameters for the task-releasing events the number $n_i$ of events and a time-window length $W_i$ in the sense that the system should still respond to up to $n_i$ such events within any sliding window of length $W_i$ in case the environment leaves the assumed envelope and issues task-$\tau_i$-releasing events more frequently than once every $T_i$.
The fraction $n_i / W_i$ gives us a rate and upper bound within which we still consider event-internalizing sensors as correct and the environment out of the anticipated envelope.
Beyond this bound, that is, when the $n_i^{th}$ event occurs within $W_i$, we raise as additional alarm, indicating that the event-triggering sensor may be faulty. It is then up to the system to decide how to respond.

Once a task $\tau_i$ releasing event occurs more frequently than once within $T_i$, feasibility changes and requires handling up to $n_i$ events within $W_i$ provided more important events and the tasks $\tau_i$ they release can still be scheduled. Like mixed-criticality feasibility, this requirement makes no claims about less important tasks, but of course we would like to maintain as many tasks as possible, ordered by importance, and ideally all of them. We therefore add as additional constraint that:\\

\begin{definition}[Out-of-Envelope Feasibility]
  At any time, system responses for up to $n_i$ task-$\tau_i$-releasing events in every sliding window of length $W_i$ must be considered and if the response is to release all instances of $\tau_i$, then those instances must be guaranteed to receive $C_i$ time between their release and deadline before any less important task $\tau_j$ (i.e., $I_j < I_i$) is scheduled.
\end{definition}
\vspace{3mm}

Note that the formulation so far also allows only responding in exceptional situations, by considering that the normal occurrence of a task has already occurred and setting this task's period to infinity.  

From here, three questions need to be answered, namely
\begin{itemize}
\item How should the system respond to more than one releasing event per $T_i$?, 
\item How to schedule released tasks so as to guarantee out-of-envelope feasibility? and
\item How to enforce that for all tasks at most $n_i$ events are handled within $W_i$, respecting the importance $I_i$ and that the system recognizes important alarms to make the appropriate adjustments?
\end{itemize}

In this preliminary work, we shall focus on the third question, but let us summarize some early conclusions about the first two questions as well.

\begin{figure}
  \begin{center}
    \includegraphics[width=.7\columnwidth]{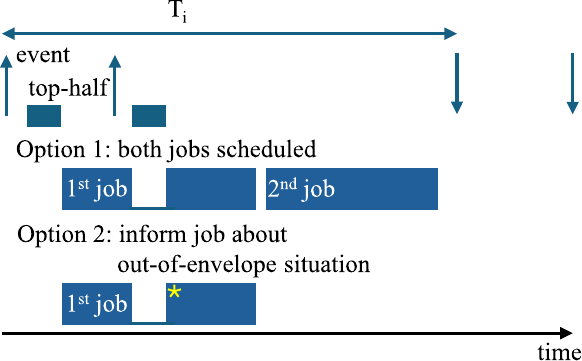}
    \caption{Internalization and scheduling options for a task, which deviates from the assumed envelope. Both events need to be internalized and generate top-half overheads. Depending on the kind of task, out of envelope behavior can be addressed by releasing all jobs for this event (option 1) or by informing an already running job (option 2), e.g., by invoking an exception handler in this task.}
    \label{fig:response}
  \end{center}
\end{figure}

Of course, we cannot avoid internalizing all $n_i$ events, which creates $n_i$-fold top-half load within the sliding window of $W_i$. However, our formulation of out-of-envelope feasibility does not require releasing all $n_i$ tasks, although for some situations this may be desirable. Alternatively, one could release some of these tasks, informing them about the out-of-envelope behavior, which then may result in the tasks triggering a criticality change to properly respond to out-of-envelope situations. Figure~\ref{fig:response} illustrates these two situations.

Clearly, out-of-envelope feasibility shows similarities to mixed-criticality feasibility, and we expect some of the results to carry over immediately.
For example, drawing inspiration from criticality-monotonic scheduling~\cite{cm}, fixed task priority assignments according to the importance of tasks will guarantee that more important tasks are scheduled before less important tasks will be considered, even in out-of-envelope situations, provided of course that under this priority assignment the task set remains schedulable in normal situations. 

\begin{figure}
  \begin{center}
    \includegraphics[width=\columnwidth]{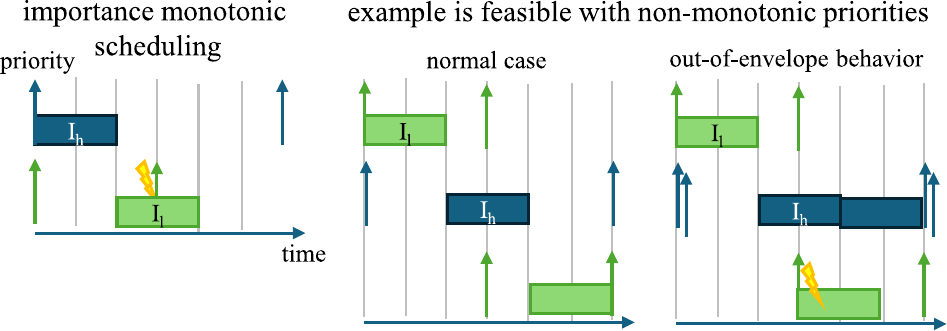}
    \caption{Example showing that importance-monotonic scheduling is not optimal. The low-importance task $\tau_l$ misses its deadline, even in normal situations (left), whereas both jobs of it can meet their deadline if the first one is higher prioritized than the high-importance task $\tau_h$ (middle). Even if $\tau_h$ is released outside the anticipated envelope (up to $n_h = 2$ events within $W_h = T_h$ in this example), $\tau_h$ meets all deadlines, at the cost of the less important second job of $\tau_l$.}
    \label{fig:optimality}
  \end{center}
\end{figure}

It is also relatively easy to construct a counterexample, showing that such ``importance-monotonic scheduling'', when releasing all tasks instances upon their releasing event, cannot be optimal. Consider the two tasks shown in Figure~\ref{fig:optimality}. The first is characterized as $\tau_l = (2, 3, I_l)$ and the second as $\tau_h = (2, 6, I_h)$, with importance $I_l < I_h$. For $\tau_h$, we consider an out-of-envelope behavior of $n_h = 2$ in every sliding window of length $W_h = T_h = 6$. Clearly, releasing the task twice is an exceptional situation, which however, we would still like to handle (e.g., to respond to an unanticipated important alarm sent in short succession). As shown in the figure, importance-monotonic priority assignment will lead to a deadline miss, even in normal situations (left). However, if we would raise the priority of $\tau_l$'s first job above the priority of $\tau_h$, we obtain a schedulable task set in normal situations (middle), but also in the exceptional situation, where $\tau_h$ is released twice (right). Remember, in the latter case we can sacrifice $\tau_l$'s second job, since it is less important than $\tau_h$. The system remains responsive to the unanticipated occurrence of the second alarm, but cannot retain the full service it had in normal situations. Also, an alarm is raised indicating that the sensor triggering this alarm operates at its bound and should be interpreted with care.

\section{Defending against Overwhelming Out-of-Envelope Behavior}
\label{sec:defense}
Ensuring that event-triggered systems will not get overwhelmed by out-of-envelope behavior requires limiting the internalization of task-releasing events to at most $n_i$ within any sliding window of length $W_i$, while providing the operating system with the signal it needs to adjust the task schedule. 
In particular, if a task is released more than once every $T_i$, this already constitutes out-of-envelope behavior and may result in sacrificing the internalization of less important events and possibly the tasks they release. This is to ensure more important events, specifically alarms, continue to be processed.

\subsection{Vectored Interrupt Controllers}
\label{sec:vics}

In this section, we illustrate how vectored interrupt controllers, such as the Advanced Programmable Interrupt Controller (APIC) variants of x86 processors or ARM's Nested Vectored Interrupt Controller (NVIC) can be configured to provide these indicators while protecting from interrupt storms.

Vectored interrupt controllers (VICs) multiplex the processor's interrupt mechanism by providing multiple external interrupt lines. In addition, they often allow interrupt lines to be assigned different priorities. In this case, the interrupt controller features an interrupt-priority level such that lower-priority interrupts can be masked simply by raising this priority level above the priority of the line. 

In this work, we assume VICs to provide such interrupt priorities and an interrupt priority level (both ARM's NVIC and various x86's APIC variants do). 
We further assume that interrupts can be masked individually and, for simplicity, will not consider sharing interrupt lines among multiple interrupt sources.
Moreover, we shall assume that the timer interrupt can be configured to be independent of the above constraints. This is possible, for example, by giving the timer the highest interrupt priority. Lastly, we shall assume devices or external interrupt controllers (e.g., IO-APIC) expose a counter per interrupt line that is incremented for each event occurring. Capture compare units already provide various such counters, albeit not for this purpose.

In the following, we shall continue to talk about priorities with the implication that these are the interrupt priorities considered in the VIC. 
We do not require interrupt priorities to correlate to task priorities and will divert from the recommendation to set the interrupt priority level to the priority of the currently running task.
Instead, we shall leverage this mechanism to control importance and prevent too many undesired interruptions.


\subsection{Detecting Out-of-Envelope Behavior}
\label{sec:more_than_once_every_T_i}

If a task were to be released more than $n_i$ times within a sliding window of size $W_i$, the releasing sensor could be considered faulty and no further events would need to be handled and, for that matter, be internalized. 

To capture out-of-envelope behavior precisely, we would have to allow all events to happen and be internalized so as to record them in ring buffers of size $n_i$, measure their time of occurrence and compute when the event leaves the task's sliding window of length $W_i$. 

\subsection{Defending Against Being Overwhelmed}

We mask interrupts to defend against being overwhelmed, which implies not internalizing such events.

In case the buffer is full, the OS masks the interrupt to prevent all subsequent occurrences of this event from being internalized, raises an alarm and sets a timer to the time of the earliest event in the buffer plus $W_i$. Being masked, subsequent events will not be internalized but are still recorded in the devices' counters. Once the timer fires, the OS unmasks the interrupt line of the event to consider further occurrences. 

At this point in time, the OS compares the counter at the device against the value it read when masking the event to identify whether the sensor is faulty (i.e., more than $n_i$ events occurred in that time). It is then up to the OS whether it will consider this sensor permanently damaged or resume using this sensor once the alarm rate drops below $n_i / W_i$.

Unfortunately, for sensors that still operate within the bound of $n_i$ occurrences within $W_i$, capturing out-of-envelope behavior precisely comes at significant costs in terms of top-half overheads and hence the interference that top-half handling causes on task execution.
In the worst case, this top-half handling, even if we capture the timestamp of the respective next event through a capture unit, amounts to $\underset{\tau_i \in \Gamma}{\Sigma} n_i \Delta_{TH}$, where $\Delta_{TH}$ denotes the time needed to enter the kernel, record the timestamp of the event in the ringbuffer, take the decision to and mask it, and returning from the kernel, since all $n_i$ occurrences of every task-releasing event may occur at once.

Obviously, this is not very feasible.
We therefore over-approximate subsequent events as if they occur with the first, in case they do not affect the current scheduling decision.
That is, we ignore all but the next occurrence of each event until when the bottom half is processed, by masking the specific interrupt in the top half and unmasking it only after the bottom half finishes executing.
Still, multiple events may occur and the device indicates their number, but we assign all of them the timestamp of the event that masked them. This reduces the sliding window during which we do not allow further events to be internalized and creates more pressure in the system\footnote{Considering events to happen earlier means the system will, at the brink of the sensor being faulty, sooner unmask interrupts, since the sliding window ends earlier.}, but ensures we do not miss events that we should have handled. 

\begin{figure}
    \begin{center}
        \includegraphics[width=1\columnwidth]{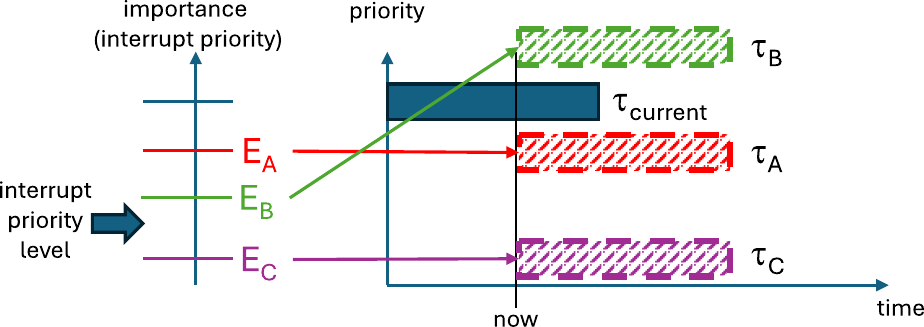}
        \caption{Example showing how the interrupt priority level may prevent internalizing some of the events that have no consequence for scheduling. In the figure, three tasks ($\tau_A, \tau_B, \tau_C$) and their events ($E_A, E_B, E_C$) are shown that may preempt the currently running task $\tau_{\mathit{current}}$. We set the interrupt priority level to below the importance of the least important task ($\tau_B$) that may still preempt $\tau_{\mathit{current}}$. This way, $\tau_C$, whose event importance $I_C$ is lower than or equal to the interrupt priority level gets masked. However, the event $E_A$ of task $\tau_A$ remains enabled, since $\tau_A$ is more important than $\tau_B$ (although less important than $\tau_{\mathit{current}}$).}
        \label{fig:importance}
    \end{center}
\end{figure}

In the same way, we mask events that have no immediate consequence on the scheduling decision.
That is, we raise the interrupt priority level to prevent internalizing the events of tasks that would not preempt the current task. These are lower-important tasks whose next job the scheduler will prioritize higher than the priority of the current running task.
Again, when lowering this level, we internalize these events with the timestamp when we masked them and evaluate the device event counts to identify whether sensor values were off.
Notice that even though this timestamp may be before the actual release time of the task, the task will not be executed before the point in time when its releasing event is internalized.
Notice also that the above mechanism is not perfect, because there may be lower prioritized and less important tasks than the currently running task, which are still more important than the task with the minimum importance whose next job will be higher prioritized. Figure~\ref{fig:importance} illustrates this point. It would of course be possible to mask these tasks manually, albeit at significant costs, in particular if the interrupt controller does not allow installing and changing interrupt masks for multiple interrupt lines at once.

\section{Limitations and Open Questions}
\label{sec:limitations}

Of course, we are not yet at the end of our journey and several limitations and criticisms remain valid, which raise open questions and require further research. 

For example, one could argue that our differentiation of the still to be considered out-of-envelope behavior and the bounds after which we consider sensors to wrongly produce events is artificial in the sense that we anticipate as out-of-envelope behavior what should actually be anticipated. 
We admit this contradiction and lack of a better definition of what out-of-envelope behavior actually is, but hope the intuition will be clear. Out-of-envelope behavior is what we did not anticipate for the normal behavior of the system but which we still might be afraid to see and therefore plan for.

Focusing on enforcement, we could of course only sketch some of the scheduling problems that occur when the system should remain feasible in out-of-envelope situations. We therefore leave as an open question, the investigation of scheduling algorithms that can guarantee schedulability of the most important tasks when inter-separation constraints are violated.

Our approach so far completely gives up on inter-separation constraints once the environment moves out of the anticipated envelope. This is mainly to capture unanticipated alarms. However, if arguments about the physical processes demonstrate that it will be infeasible that events occur more frequently than once every $T^{\mathit{out}}_i$, then integrating such bounds in the task model would greatly improve the schedulable utilization of the system, even if the out-of-envelope period $T^{\mathit{out}}_i$ is small compared to the period $T_i$.

Likewise, our approach so far assumes importance to be a total order among tasks. It might be interesting to investigate more well-defined structures that can capture different sets of tasks that should be sacrificed depending on which event violates the assumptions made in the system's operational envelope. 

Being a report on preliminary work, we obviously did not quantify the overhead of our approach on existing vectored interrupt controllers, nor did we evaluate the design of VICs that are specifically designed to capture out-of-envelope behavior. For example, the constraints which prevented us from precisely tracking out-of-envelope events in Section~\ref{sec:defense} vanish if ringbuffers containing timestamps are maintained in the VIC. Like capture units, such VICs could record the timestamp of occurrence of interrupt-triggering events (provided the $n_i$-element ringbuffer is not yet full) while triggering the processor's interrupt service routine only when the processor is ready to receive such an interrupt and when it would affect scheduling decisions (communicated by the masking scheme we discussed above).

\section{Related Work}
\label{sec:related_work}

In this work, we have already discussed similarities but also differences to mixed-criticality scheduling~\cite{4408308}, whose state of the art is captured in the review by Davis and Burns~\cite{burns2013mixed}. Importance allows us to define trading off tasks independent of those that originate from high-criticality tasks exceeding their low-criticality expectations and we hope its use to respond to clearly unanticipated situations avoids some of the misconceptions of mixed-criticality systems~\cite{7527646}.

We already mentioned several works to investigate event-triggered systems and what properties they may retain despite giving up on internalizing events only according to the globally synchronized sparse time base.
For example, Scheler and Schroeder-Preikschat~\cite{5755437} ask whether the difference between event- and time-triggered is just a matter of configuration.
We believe this can be answered only after extending our observations to the network level.

We likewise mentioned that others have also proposed solutions for defending against interrupt storms (with the purpose of remaining within the operational envelope, unlike what we propose — to slightly step out of it and focus on the most important tasks).
For example,
Parmer and West~\cite{4700438} describe a means for predictable interrupt management with the help of deferrable servers, which, among others, has enabled core-local reasoning and predictable cross-core communication~\cite{10568040} in the M3 microkernel~\cite{m3}.
Scheler et al.~\cite{10.1145/1629395.1629419} discuss hardware supported interrupt handling for event-triggered real-time operating systems.
Kim et al.~\cite{7299849} explore interrupt handling and enforcement in real-time system virtualization.
Ley va-del Foyo et al.~\cite{10.1145/2220336.2220344} propose to integrate interrupt and task handling.
Brinkschulte et al.~\cite{article} suggest in a work-in-progress presentation the notion of interrupt service threads and Elliott and Anderson~\cite{6257578} propose robust interrupt handling for multiprocessor systems, by drawing inspiration from GPUs.

\section{Conclusions}
\label{sec:conclusions}

In this work, we raise the question whether real-time systems can still remain responsive in case unanticipated and unforeseen combinations of events move them beyond the operational envelope for which they have been designed. We distinguish normal situations from situations where such out-of-envelope behavior occurs and from situations where faulty sensors overwhelm the system by generating interrupt storms or event combinations that the system can no longer sensibly handle. We propose a vectored interrupt controller-based event internalization and handling scheme that is capable of defending against the latter, while supporting the two former.
Our scheme can be implemented using existing interrupt controllers (e.g., on ARM and x86), but would at the same time greatly benefit from a dedicated capture unit that records the timestamps of up to $n_i$ interrupt-triggering events in per interrupt-line ringbuffers.

Directions for our future work include evaluating such hard- and software implementations, further exploring the similarities and differences between importance-based scheduling (once the system leaves its operational envelope) and mixed-criticality scheduling, and investigating whether out-of-envelope behavior can also be tolerated at the network level.

\section{Acknowledgments}
The authors would like to thank the anonymous reviewers for their helpful comments.

\IEEEtriggeratref{12}
\bibliographystyle{IEEEtran}
\bibliography{refs}


\end{document}